\documentclass[]{elsart}
\usepackage{epsfig,longtable,graphicx,subfigure,color}
\begin{document}
  \begin{frontmatter}
    \title{The ZEPLIN II dark matter detector: data acquisition system and 
      data reduction}
    \author[ral]{G.J. Alner}, \author[ic,ral]{H.M. Ara\'{u}jo}, 
    \author[ic]{A. Bewick},
    \author[ral,ic]{C. Bungau}, \author[ral] {B. Camanzi}, 
    \author[shef]{M.J. Carson\corauthref{cor}},
    \corauth[cor]{Corresponding author.}
    \ead{m.j.carson@sheffield.ac.uk}
    \author[shef]{H. Chagani},
    \author[coimbra]{V. Chepel}, \author[ucla]{D. Cline},
    \author[ic]{D. Davidge}, \author[shef]{J.C. Davies}, \author[shef]{E. Daw},
    \author[ic]{J. Dawson}, \author[ral]{T. Durkin}, \author[ic,ral]{B. Edwards},
    \author[shef]{T. Gamble}, \author[ucla]{J. Gao},\author[edin]{C. Ghag},
    \author[ic]{W.G Jones}, \author[ic]{M. Joshi}, \author[edin]{E.V. Korolkova},
    \author[shef]{V.A. Kudryavtsev}, \author[shef]{T. Lawson},
    \author[ic]{V.N. Lebedenko},
    \author[ral]{J.D. Lewin}, \author[shef]{P.K. Lightfoot},
    \author[coimbra]{A. Lindote}, \author[ic]{I. Liubarsky}, 
    \author[coimbra]{M.I. Lopes},
    \author[ral]{R.L\"{u}scher},
    \author[shef]{P. Majewski}, \author[shef]{K. Mavrokoridis},
    \author[shef]{J.E. McMillan}, \author[shef]{B. Morgan},
    \author[shef]{D. Muna}, \author[edin]{A.S. Murphy},
    \author[coimbra]{F. Neves},
    \author[shef]{G. G. Nicklin}, \author[ucla]{W. Ooi},
    \author[shef]{S.M. Paling}, \author[coimbra]{J. Pinto da Cunha},
    \author[edin]{S.J.S. Plank}, \author[ral]{R.M. Preece},
    \author[ic]{J. Quenby},
    \author[shef]{M. Robinson}, \author[coimbra]{C. Silva}, 
    \author[coimbra]{V.N. Solovov},
    \author[ral]{N.J.T. Smith},
    \author[ral]{P.F. Smith}, 
    \author[shef]{N.J.C. Spooner},
    \author[ic]{T.J. Sumner}, 
    \author[ic]{C. Thorne}, \author[shef]{D. Tovey}, \author[shef]{E. Tziaferi},
    \author[ic]{R.J. Walker}, \author[ucla]{H. Wang}, \author[texas]{J. White},
    \author[roch]{F.L.H. Wolfs}
    \address[ral]{Particle Physics Department, Rutherford Appleton
      Laboratory, Chilton, UK}
    \address[ic]{Blackett Laboratory, Imperial College London, UK}
    \address[shef]{Department of Physics \& Astronomy, University of
      Sheffield}
    \address[coimbra]{LIP-Coimbra \& Department of Physics, University
      of Coimbra, Portugal}
    \address[ucla]{Department of Physics \& Astronomy, University of
      California, USA}
    \address[edin]{School of Physics, University of Edinburgh, UK}
    \address[texas]{Department of Physics, Texas A\&M University, USA}
    \address[roch]{Department of Physics \& Astronomy, University of
      Rochester, USA} 
    \begin{abstract}
      ZEPLIN\,II is a two-phase (liquid/gas) xenon dark matter
      detector searching for WIMP-nucleon interactions. In this paper
      we describe the data acquisition system used to record the 
      data from ZEPLIN\,II and the reduction procedures
      which parameterise the data for subsequent analysis.
    \end{abstract}
  \end{frontmatter}
  \section{Introduction}
  The ZEPLIN\,II dark matter detector has been operational at the
  Boulby Mine underground laboratory since 2005. Its principle aim is
  to detect and measure the faint nuclear recoil signal from galactic
  Weakly Interacting Massive Particles (WIMPs). ZEPLIN\,II is a
  two-phase (liquid/gas) xenon detector which has an increased
  sensitivity over previous UK Dark Matter Collaboration experiments
  NaIAD~\cite{naiad} and ZEPLIN\,I~\cite{zeplin1}. By measuring both
  the primary and secondary scintillation signals produced by
  particles interacting in the target volume~\cite{davies,wang,cline}
  this technique has the potential to improve discrimination between
  nuclear recoils expected from WIMP interactions (also produced in
  neutron collisions with nuclei) and electron recoils (caused by
  $\gamma$-rays and electron interactions, e.g. $\beta$-particles).
  The differential energy spectrum of these nuclear recoil events is
  expected to be featureless and smoothly decreasing with detected
  recoil energies which are less than 100\,keV for WIMP masses in the
  range 10-1000\,GeV\,c$^2$~\cite{lewinsmith}. Dark matter experiments
  need to be capable of detecting recoil energies of a few keV in
  order to place the most stringent limits on the rate of WIMP
  interactions. In scintillator experiments this requires
  sensitivities to single photoelectrons. Due to the extended waveform
  digitisation and fine resolution, a large volume of data is collected
  ($\sim$8 TBytes/yr, excluding calibration data). The data requires
  efficient processing and must be stored for subsequent
  analysis. Reduction procedures capable of parameterising the data
  have been developed to process the waveforms and output a set of
  parameters representative of the original waveform.

  In this paper we present a brief description of the detector
  followed by a more detailed discussion of the data acquisition
  system and data reduction procedures. In our
  companion paper~\cite{z2science} we present initial results from the
  first underground run of ZEPLIN\,II.
  \section{The Detector}
\begin{figure}
    \begin{center}
      \includegraphics[width=20pc]{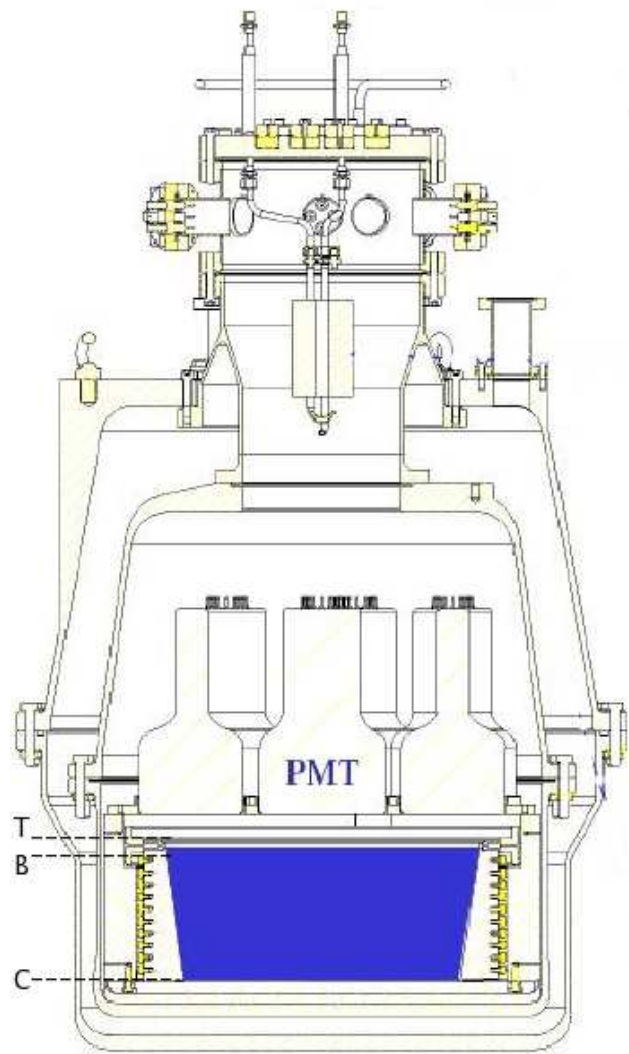}
      \caption{Schematic of ZEPLIN II showing the relative positions
	of the electric field grids and the liquid/gas interface where T
	is top grid, B is bottom grid and C is the cathode.}
      \label{z2pic}
    \end{center}
  \end{figure}

  ZEPLIN\,II consists of 31\,kg of liquid xenon contained in a 50\,cm
  diameter copper vessel (Figure~\ref{z2pic}). The target volume is
  viewed from above by 7 ETL D742QKFLB 130\,mm photomultiplier tubes
  (PMTs) in a close-packed hexagonal pattern. Two steel grids and a
  mesh act as electrodes which define the electric field in the
  detector. Two of the grids are positioned above and
  below the liquid surface; these are the top and bottom grids which
  define the electroluminescence and charge extraction fields. The
  wire mesh is located at the bottom of the target defining the drift
  field in the active volume of the target. PTFE lining the inside of
  the copper vessel acts both as a support structure for the high
  voltage grids and a reflector. The target is maintained at liquid
  xenon temperature by an IGC PFC330 Polycold
  system~\cite{polycold}. Circulation of the xenon through SAES
  getters (model PS11-MC500)~\cite{saes} ensures impurities do not
  reduce the performance of the target. The copper vessel is
  surrounded by a stainless steel jacket which provides an
  insulating vacuum. The detector sits in a liquid scintillator veto
  system viewed from above by 10 ETL 9354KA 200\,mm PMTs. The upper
  half of detector/veto system is surrounded by hydrocarbon slabs
  (separated by Gd-loaded resin sheets) 
  shielding the target from external background neutrons (the liquid
  scintillator veto acts as its own neutron shield). The entire
  apparatus is housed in a lead ``castle'' which shields against
  external background $\gamma$-rays.

  In normal (two-phase) operating mode ZEPLIN\,II is designed to
  detect the vacuum ultra-violet scintillation and ionisation charge
  signal of an interacting particle. The scintillation light comes
  from the initial particle interaction (prompt de-excitation of
  Xe$_2^*$ dimer to dissociative ground state).  Under an applied
  electric field, a fraction of the ionisation electrons are drifted
  to the liquid surface where they are removed by the extraction field
  into the gas phase producing a secondary scintillation pulse through
  electroluminescence. We adopt the convention of naming the primary
  and secondary signals S1 and S2 respectively.

  The region between the bottom grid and the cathode defines the
  active volume of the detector and contains a drift field of
  $\sim$1\,kV/cm. The electron drift velocity in this region is about
  2.0\,mm/$\mu$s. Field shaping rings embedded in the PTFE support
  structure keep the drift field lines parallel. The top grid defines
  both the extraction and electroluminescence fields in the region
  between the top and bottom grids. The time delay between the S1 and
  S2 signals permits the depth of the particle interaction to be
  reconstructed.

  \section{Data Acquisition System}
\begin{figure}
    \begin{center}
      \includegraphics[width=35pc]{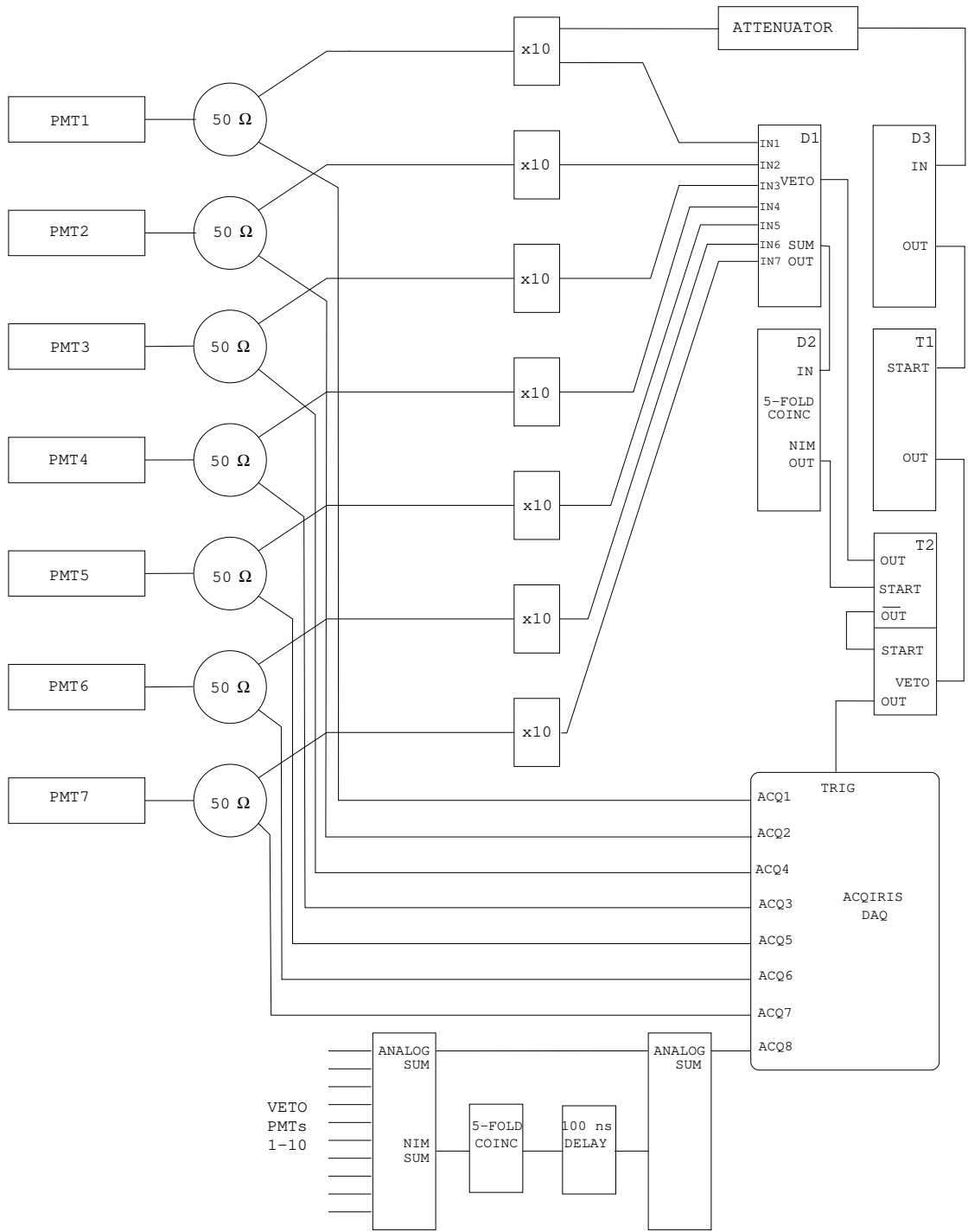}
      \caption{Signal path from target and veto PMTs to DAQ via trigger
	electronics. }
      \label{fig:trigger}
    \end{center}
  \end{figure}

  Figure~\ref{fig:trigger} shows the signal path from PMT to the data
  acquisition system (DAQ).  PMT
  gains are equalised to give about 4.5\,mV per photoelectron (pe)
  output. The 7 PMT signals from the target are first passively split by a
  50\,$\Omega$ splitter: one line is fed to a $\times$10 amplifier,
  the other to the signal input of the DAQ (channels ACQ1 to ACQ7 in
  Figure~\ref{fig:trigger}). From the amplifier the signals are fed to
  discriminator D1 which outputs 50\,mV/channel for input signals
  above 17\,mV (approximately 2/5 of the spe signal).  The logic sum
  of the signals is then fed to discriminator D2 which outputs a NIM
  pulse to dual timer T2 when 5 out of 7 PMTs detect a signal above
  the 17\,mV threshold level. At the same time, T2 sends a 100\,$\mu$s
  square-wave pulse to the veto input of D1 to prevent further
  triggers until the whole waveform is read by the DAQ.  The amplified
  signal from the central PMT (PMT 1) is also attenuated
  before going into discriminator D3. Due to the relative placing of
  the PMTs and the PTFE support/reflector structure, the central PMT
  sees a larger signal (on average) than the outer PMTs. Large
  amplitude signals can cause optical feedback in the target giving
  rise to many noise pulses of long duration. Signals exceeding a
  pre-set threshold of 200\,mV (see below for explanation) are vetoed
  by the dual timer T2 which sends a 1\,ms inhibit signal to D1 to
  prevent optical feedback signals from triggering the system. This
  also has the effect of reducing the trigger rate by 60\%, reducing
  data processing and data storage requirements. The 10 PMT signals
  from the veto feed 10 channels of a discriminator/buffer NIM
  module. The sum of the discriminator outputs is passed into a second
  discriminator whose threshold is set to output a logic pulse if at least 3
  of the veto PMTs fire in coincidence. This pulse is delayed by
  100\,ns in a delay line and the output is added to the analog sum of
  the veto PMT outputs.
  
  The waveform hardware consists of DC265 M2M ACQIRIS digitizers embedded
  in CC103 ACQIRIS~\cite{acqiris} crates. These are based on
  CompactPCI technology interfaced through the PCI bus. The digital
  conversion of signals has an 8\,bit resolution, a conversion rate of
  up to 500\,MSamples/s, a bandwidth of 150\,MHz and a memory of
  2\,MPoints/channel. Each 200\,$\mu$s waveform is sampled at 2\,ns
  intervals. LINUX-based software reads out the digitised waveforms
  which are then written to disc. Monitoring of all target parameters
  such as temperature and pressure is done with a 64 channel
  Datascan~\cite{datascan} module via the serial port on the DAQ
  computer.

\begin{figure}
    \begin{center}
      \includegraphics[width=30pc]{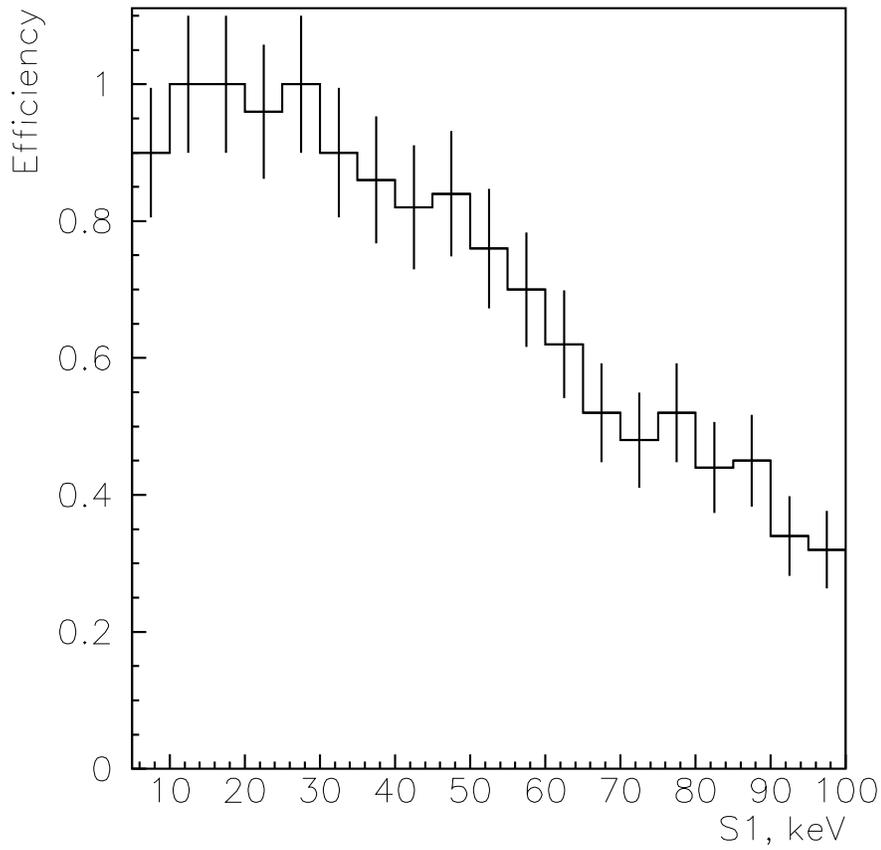}
      \caption{Software saturation cut efficiency as a function of
	energy. Efficiency is 100\% up to 30\,keV. }
      \label{daqeff}
    \end{center}
  \end{figure}

  Signals exceeding an amplitude of 200\,mV are vetoed by the DAQ
  electronics. In addition, an upper cut of 180\,mV is implemented in
  software. Signals above this threshold are above the
  energy range of interest for dark matter searches but the effect, in
  terms of efficiency, must be estimated. Figure~\ref{daqeff}
  shows the efficiency as a function of energy calculated from data
  taken with and without the software threshold - but keeping
  the additional 200\,mV saturation cut on the central PMT in both
  cases. The efficiency is 100\% up to 30\,keV.
 
  In order to investigate the fraction of events lost due to DAQ
  dead time dedicated pulser measurements were performed. The pulser
  was set to output a pulse of amplitude 0.81\,V with 0.2\,ms
  duration. For a waveform of 100,000 samples at 2\,ns per sample the
  maximum recordable rate was 22\,Hz, corresponding to a dead-time of
  $\sim$50\,ms. In a non-paralyzable model, where events occuring
  during dead periods do not extend the dead-time, the measured event
  rate $n$ is given by $n=m/(1-m\tau)$~\cite{knoll}, where $m$ is the
  observed event rate and $\tau$ is the dead-time. The typical
  observed background rate in the target is 2\,Hz which corresopnds to
  a loss of $\sim$10\% of events (since $m/n=1/(1-n\tau)$). For data
  taken with an AmBe neutron source located approximately 1\,m above
  the target the true event rate was 34.4\,Hz which corresponds to a
  loss of 67\%. For a $^{60}$Co source located in the same position
  the loss was 78\% with a true event rate of 70\,Hz.
  
  \section{Data Reduction}
  A raw data file with 2000 events is approximately 250\,MB compressed
  (each waveform has 100,000 points and there are 7 PMT channels plus
  1 veto channel for each event). Approximately 25\,GB of data
  (excluding calibration runs) is recorded each day. The data are then
  written to magnetic digital tape on an ADIC Scalar 100 tape robot
  system~\cite{adic}. Each tape can store 100\,GB of data. The data
  are subsequently transferred to an Apple XGrid capable
  of reducing $\sim$0.6\,TB of data per day.

  Raw data are reduced with a LINUX-based application
  which reads in the binary data files and
  outputs a set of numeric parameters representing each pulse found
  on each waveform. The software consists of an event viewer, allowing
  the examination of each trace in each PMT and reduction algorithms
  which process the waveforms from each channel.  All peaks in each
  waveform must be identified and parameterised according to height,
  width, area and time-constant. Pulse parameters are then written to
  reduced data files in HBOOK ntuple format~\cite{hbook} for subsequent
  analysis. The user must specify a number of input variables for the
  peak-finding algorithms, these are discussed in turn.

  \subsection{Input Parameters}
\begin{figure}
    \begin{center}
      \includegraphics[angle=0,width=30pc]{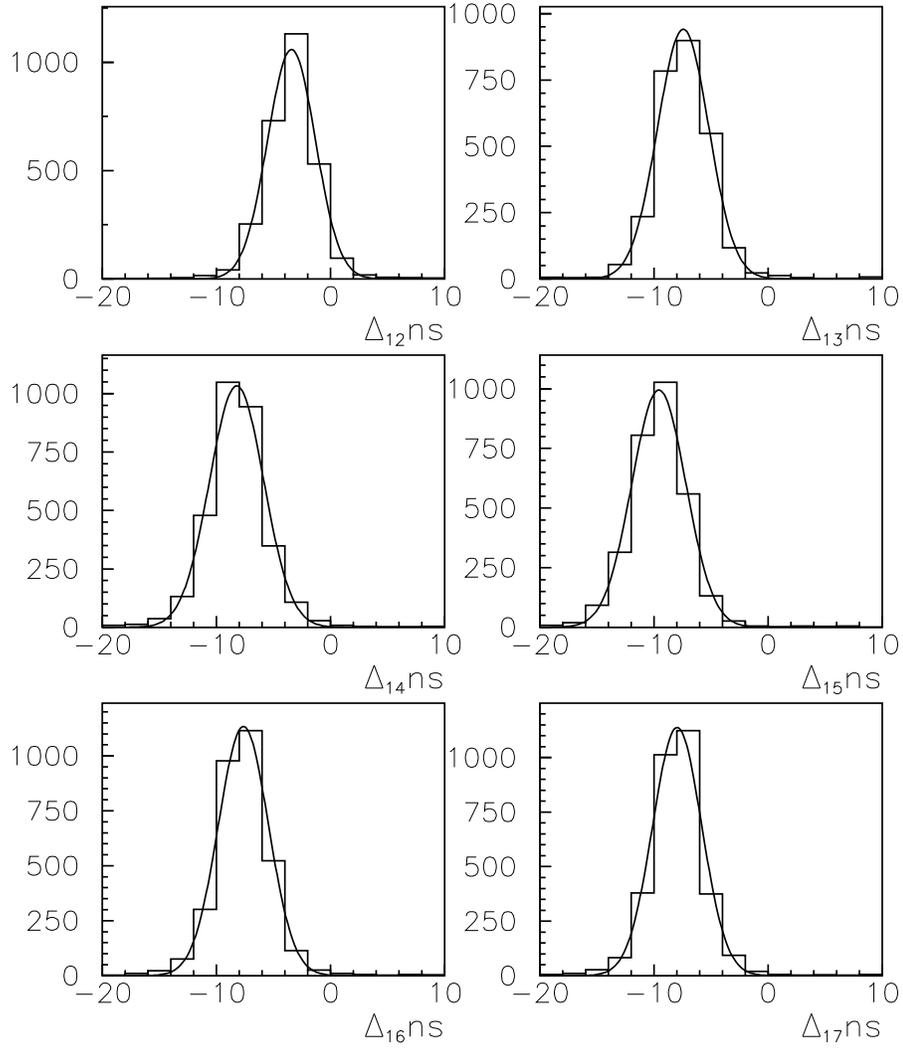}
      \caption{Mean signal delay $\Delta_{1i}$ between PMT\,1 and
	PMT\,i. The $\Delta_{1i}$ are 3.927\,ns, 7.941\,ns, 8.737\,ns,
	10.07\,ns, 6.294\,ns and 6.916\,ns for $i=2\ldots7$ respectively.}
      \label{fig:delay}
    \end{center}
  \end{figure}

  Different length signal cables from the PMTs to the DAQ and
  differing PMT characteristics can induce delays in the pulse arrival
  times in each channel. In addition, differences in transit times in
  the PMTs can induce delays up to $\sim$40\,ns. Pulses detected
  coincidently in several channels can appear spread out in the summed
  waveform if these delays are not corrected. This can lead to peaks
  in the summed waveform being incorrectly parameterised as separate
  pulses. Figure~\ref{fig:delay} shows the distribution of pulse mean
  arrival times in each PMT relative to the central PMT (PMT\,1) for
  uncorrected data.  Channels are shifted in software by their delay
  with respect to PMT\,1 and the summed waveform calculated.

\begin{figure}
    \begin{center}
      \includegraphics[angle=0,width=30pc]{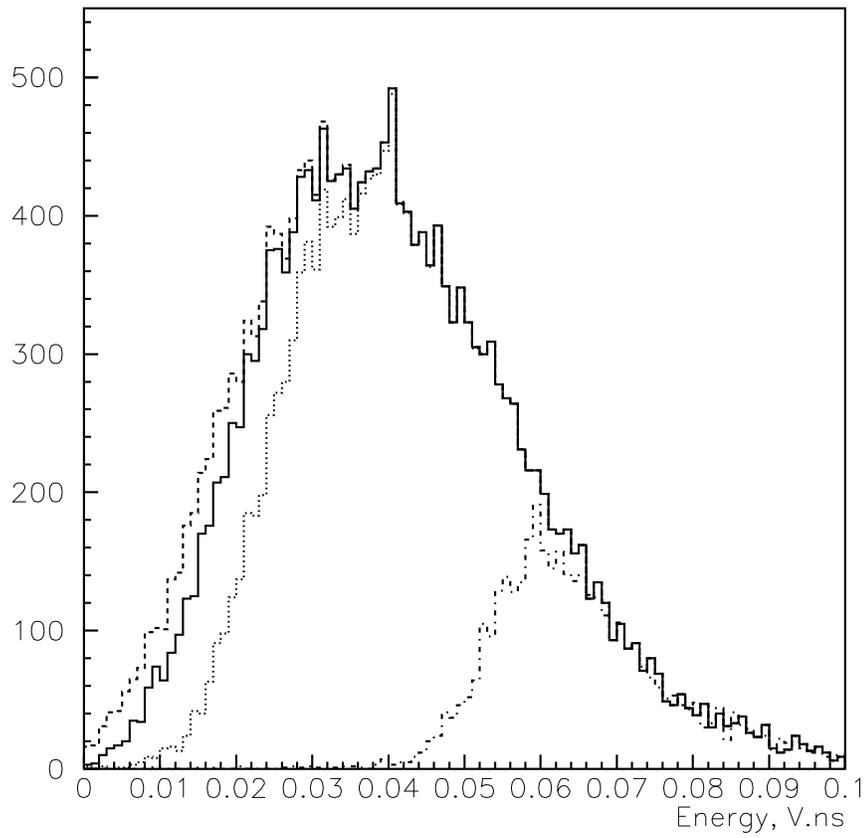}
      \caption{Single photoelectron spectra for PMT\,1 with
	different values of $t_{smooth}$: 5\,ns (dashed line), 12\,ns (solid line),
	20\,ns (dotted line) and 50\,ns (dash-dot line). The spectrum
	peaks at $\sim$0.035\,pe/keV}
      \label{fig:spes}
    \end{center}
  \end{figure}

  To facilitate peak-finding a smoothing function is applied to the
  summed waveform. This ensures that small amplitude fluctuations do
  not get mis-identified as valid signal pulses. The amplitude $h_t$
  at each sample is smoothed as
  \begin{center}
    \begin{equation}
      S_t=\frac{\sum_{t=-t_{sm}/2}^{t_{sm}/2}{h_t}}
      {N_{-t_{sm}/2}^{t_{sm}/2}}
      \label{eq:smooth}
    \end{equation}
  \end{center}
  where $t_{sm}$ is the smoothing timescale and is an input
  parameter to the reduction. Figure~\ref{fig:spes} shows a single
  photoelectron spectrum from the central PMT parameterised with different
  values of $t_{sm}$ from 5\,ns up to 50\,ns. The larger value of 
  $t_{sm}=50$\,ns can be seen to 'wash-out' lower energy pulses
  resulting in a shift in the peak of the spectrum to higher
  energies. A value of $t_{sm}=12$\,ns was chosen as it correctly
  reproduces the mean value of the spe calibrations.

  All pulses above a user-defined software threshold are tagged, up to
  a maximum of 10. The software threshold depends on the full-scale of
  the DAQ and is set to 2\,mV (almost 1/2 pe) for data aquired with a
  full scale of 200\,mV. This range was chosen as a compromise between
  energy resolution for S1 signals and the dynamic range for S2
  pulses. The smoothed amplitude $h_s$ is used only to locate peaks on
  the summed waveform, all other parameterisation uses the unsmoothed
  original data.

\begin{figure}
    \centering
    \subfigure[]{\epsfig{figure=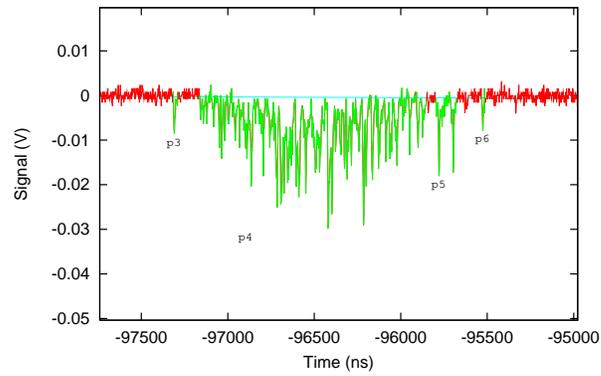,width=8.0cm}}\\
    \subfigure[]{\epsfig{figure=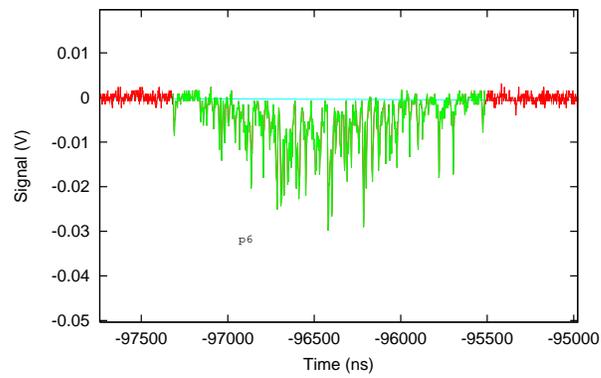,width=8.0cm}}
    \caption{Ionisation pulse parameterised with 50\,ns clustering
      (top) showing that four different pulses ({\tt p3} to {\tt p6}) have
      been identified. With 400\,ns clustering (bottom) these have
      been correctly grouped into a single pulse {\tt p6}.}
    \label{fig:cluster}
  \end{figure}

  We define a clustering timescale which allows closely spaced peaks
  on the waveform to be grouped together to form a single pulse. A low
  energy S2 signal can appear as separate peaks spread out over several
  $\mu$s (the total width of S2 is determined by the distance
  between the top grid and the liquid
  surface). Figure~\ref{fig:cluster} shows the effect of different
  clustering values on the same pulse. A clustering of 50\,ns causes
  peaks {\tt p3, p5} and {\tt p6} to be identified as separate pulses
  from {\tt p4}. Increasing the clustering timescale to 400\,ns
  correctly groups all peaks as a single pulse.

  \subsection{Output Parameters}
   The baseline for each waveform is calculated on an event-by-event
  and channel-by-channel basis.  An initial baseline is calculated
  from the mean of the first 500 data points.  However, this is not
  sufficient because the baseline tends to wander bay a few mV during
  the event. A box-car smoothing algorithm with a width of 5000\,ns is
  used to calculate a wandering baseline to compensate for this
  effect.  As the wandering baseline is calculated, any data point
  deviating by more than 5$\times$ the RMS noise from the current
  value of the baseline is excluded from the baseline calculation.
  This prevents the wandering baseline from following the slow S2
  signals from the gas phase of the data.

\begin{figure}
    \begin{center}
      \includegraphics[angle=0,width=40pc]{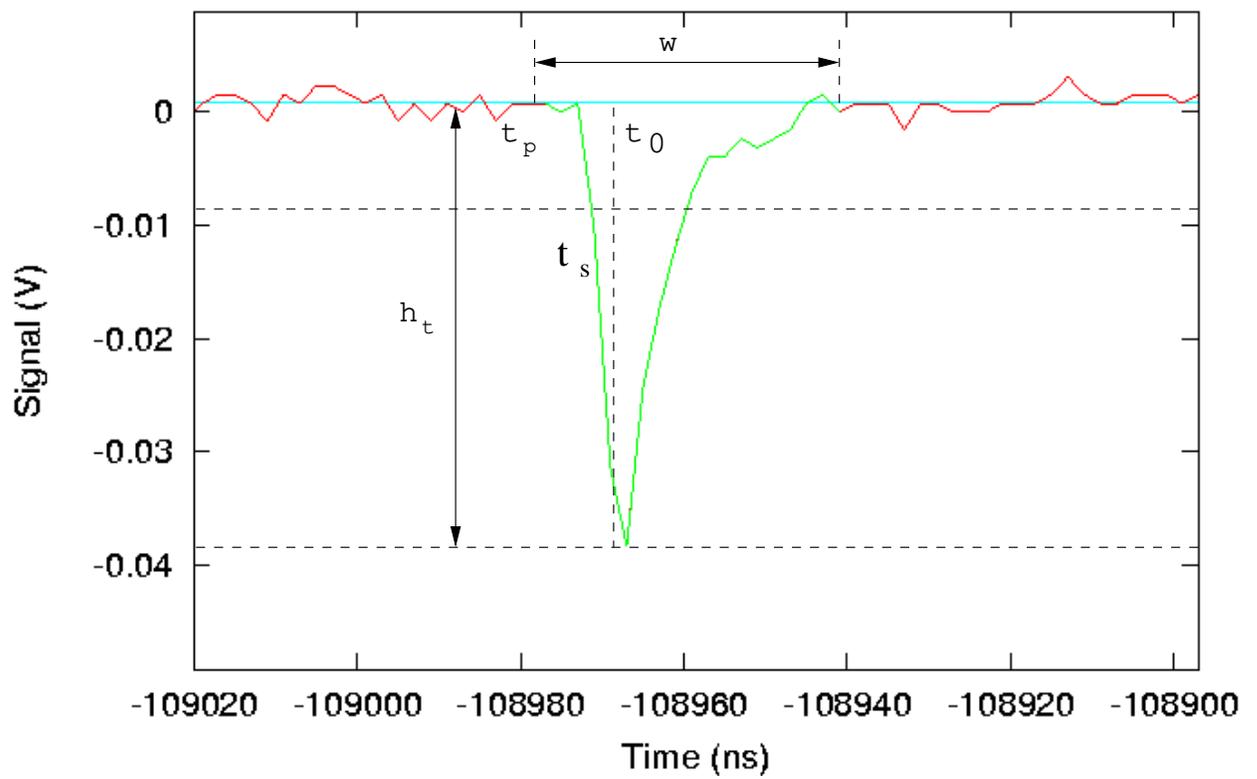}
      \caption{Basic pulse parameters. $h_t$ is the signal amplitude
	above the baseline, $t_p$ is the start time of the pulse and
	$t_0$ is the time at which 10\% of the total charge has
	arrived. $w$ is the pulse width defined by the start and end
	time of the peak.}
      \label{fig:example}
    \end{center}
  \end{figure}

  An initial scan is made of the waveform on the sum channel for
  pulses with amplitudes above the software threshold. The start
  time of each pulse $t_s(i)$ (see Figure~\ref{fig:example}) where $i$ is
  the index of the pulse (up to the maximum of 10) is then used to
  find the proper start time of the pulses $t_p(i)$ on the
  baseline. The difference between the end and start time defines the
  pulse widths $w(i)$.  Once all pulses on the summed waveform are
  identified, each individual channel is scanned in the time windows
  $w(i)$.  Each $t_p(i)$ is used to define a time $t_0(i)$, at which
  10\% of the total charge of the pulse is detected. This is then used
  to calculate the charge mean arrival time $\tau$ as in
  Equation~\ref{eq:tau}.
     
  \begin{equation}
    \tau=\frac{\sum_{t=t_0}^{t_p+w}{h_t.(t-t_0)}}{\sum_{t=t_0}^{t_p+w}{h_t}}
    \label{eq:tau}
  \end{equation}
  
  where $h_t$ is the largest amplitude within the time window. The FWHM of
  the pulse is calculated starting from the time at which $h_t$ is
  observed and tracing outwards in both directions to the times at
  which the pulse height falls to $h_t/2$. Another measure of the full
  width half maximum (LFWHM) is calculated by tracing the pulse height
  to its half maximum value by starting at both the beginning and end
  of each pulse. 

  The RMS noise of the waveform is the mean deviation of the data from
  the baseline in the pre-trigger region.
  \begin{equation}
    N_{RMS}=\sqrt{\frac{\sum{h_t}^2}{N}}
  \end{equation}

  Pulse area $A$ (in units of nV.s) is defined as the integrated area
  of the pulse in the time window and is intended as the best measure
  of the total charge associated with the pulse.
  \begin{equation}
    A=\sum_{t=t_p}^{t_p+w}{A_t}
  \end{equation}

  The summed veto signal is parameterised according to its height, arrival time
  and total integrated charge following the same procedure for signal
  pulses from the target. Table~\ref{table:parameters} lists all pulse
  parameters calculated.

  \section{Data Analysis}
  Reduced data files consist of parameters for each pulse found on
  each of the 7 PMT channels defined by those pulses found on the sum
  channel.  All parametersied waveforms must be analysed to extract
  the required S1 and S2 signals. The secondary ionisation signal will
  be delayed from the primary scintillation signal by an amount which
  depends upon the depth of the interaction in the target. The drift
  velocity of ionistation electrons (which is determined by the drift
  field) is typically 2.0\,mm/$\mu$s. The longest drift time is the
  time taken for ionisation electrons from the cathode to reach the
  bottom grid and is about 75\,$\mu$s.  Since the DAQ can trigger on
  either the S1 or S2 signal, waveforms are recorded 100\,$\mu$s on
  each side of the trigger position. The waveform time window was
  chosen to be [-200,0]\,$\mu$s with the trigger at -100\,$\mu$s.
  Events which trigger on the primary will have a secondary signal in
  the range [-100,0]\,$\mu$s and those events which trigger on the
  secondary will have a primary scintillation signal in the range
  [-200,-100]\,$\mu$s. In either case, secondary ionisation signals
  only occur in the range [-100,0]\,$\mu$s. The first step in the
  analysis involves scanning this region for pulses on the sum channel
  which are greater than 1\,V.ns in energy (see below). All pulses
  greater than 1\,V.ns in this range are tagged as possible S2
  signals. Next, the [-200,-100]\,$\mu$s region of the wavefom is
  scanned for the primary sciltillation pulse. Coincident pulses with
  an amplitude greater that 1.7\,mV in any 3 out of 7 channels are
  tagged as possible S1 candidates. An additional cut of
  $\tau>150$\,ns for S2 and  $2\,$ns\,$<\tau<50$\,ns for S1 is
  applied (the efficiencies for these and other selection criteria are
  discussed in~\cite{z2science}). We reject those events with more
  than 1 S2 signal as WIMPs do not multiple scatter. We also reject
  events with more than 1 primary scintillation pulse. All events with
  {\it only} one S1 and {\it only} one S2 are accepted.

\begin{figure}
    \begin{center}
      \includegraphics[angle=0,width=30pc]{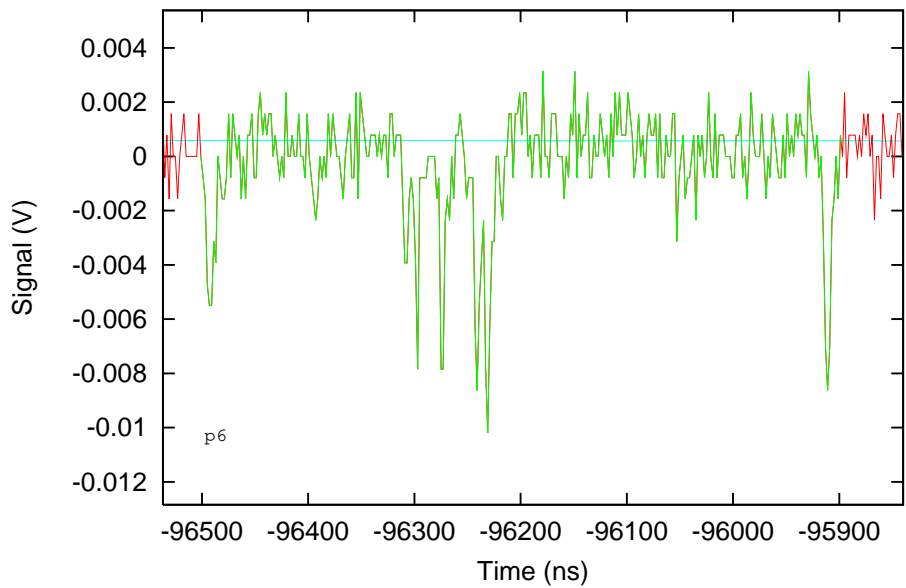}
      \caption{Typical noise event due to single electrons extracted
	from the liquid giving rise to an electroluminescence signal in the gas.}
      \label{fig:s1e1}
    \end{center}
  \end{figure}

\begin{figure}
    \begin{center}
      \includegraphics[angle=0,width=30pc]{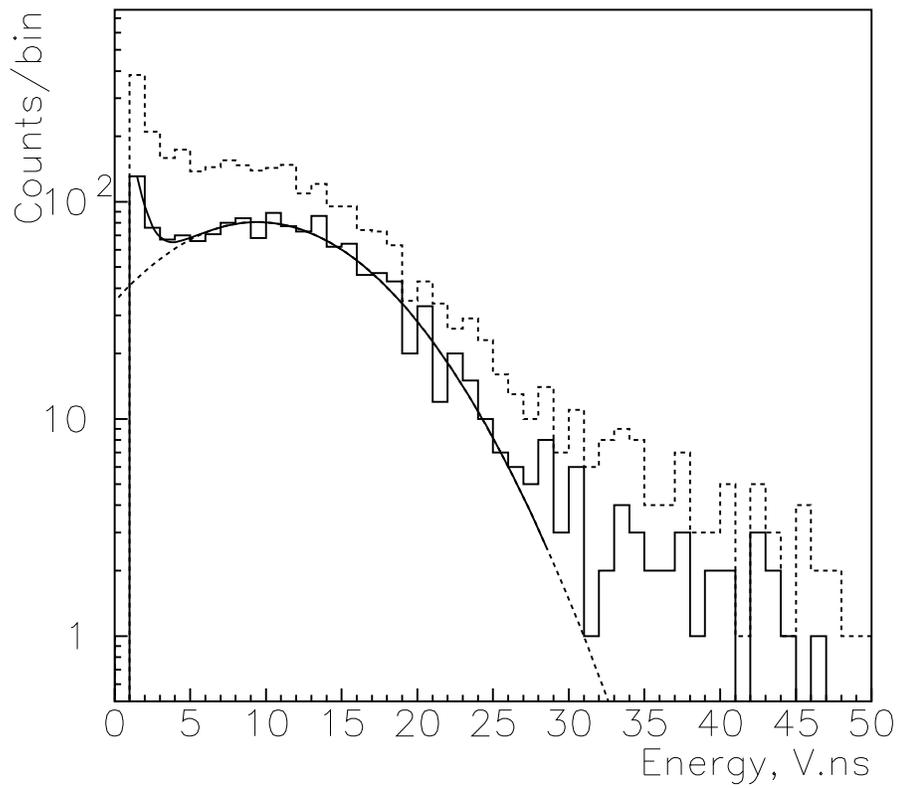}
      \caption{AmBe S2 distribution for S1 between 5\,keV and
	10\,keV (dashed histogram). The solid histogram is the
	distribution excluding events near the top and bottom
	grids. Superimposed is the Gaussian fit (plus exponential noise)
	to the data.}
      \label{fig:s2}
    \end{center}
  \end{figure}

  The 1\,V.ns selection criteria ensures that small single electron
  signals do not get mis-identified as S2. Figure~\ref{fig:s1e1} shows
  a typical cluster of single electron pulses spread over 600\,ns with
  a total energy of 0.5\,V.ns. Each peak appears in different channels
  at different times (not shown) but the clustering parameter groups
  them together in a single pulse. Low energy nuclear recoils are
  expected to produce more than one ionisation electron so the
  probability of rejecting genuine S2 signals is low. To investigate
  this further, we plot the distribution of S2 for S1 in a restricted
  energy range in Figure~\ref{fig:s2}. The distribution peaks at
  $\sim$10\,V.ns and is roughly Gaussian in shape with a noise
  contribution impinging on the distribution from the left. Fitting an
  exponential plus Gaussian to the S2 spectrum allows us to calculate
  the efficiency of the 1\,V.ns cut by integrating the area of the
  Gaussian above 1\,V.ns and dividing by the total area. Requiring a
  minimum of 1\,V.ns for S2 results in loss of approximtely 8\% of
  events (92\% efficiency) between 5-10\,keV. However, the fitted
  Gausian extends to below zero, which is unphysical, and so this loss
  is overestimated. For higher energies the efficiency is close to
  100\%.

\begin{figure}
    \begin{center}
      \includegraphics[angle=0,width=30pc]{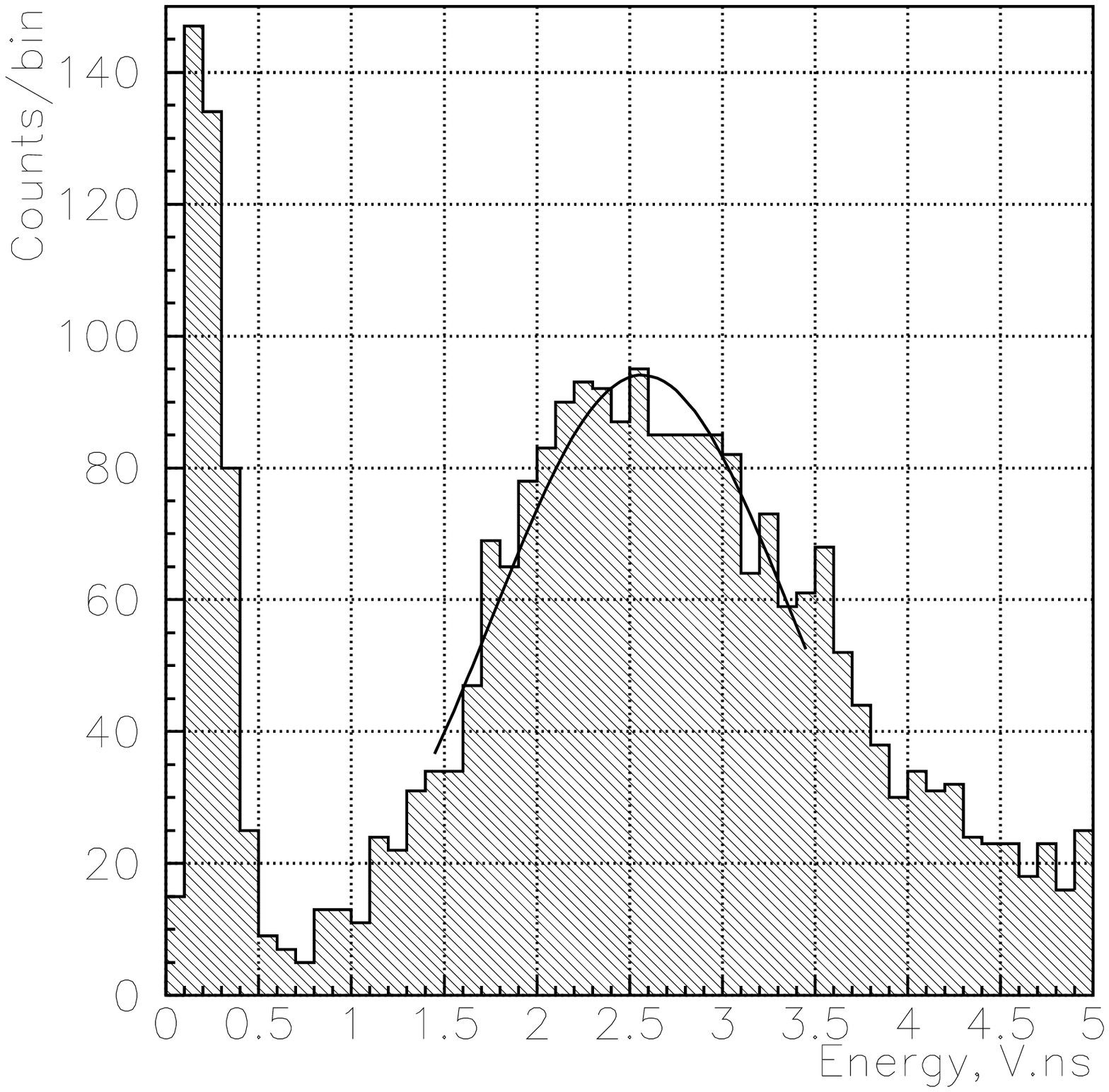}
      \caption{Energy spectrum from $^{57}$Co calibration run with a
	drift field of $\sim$1\,kV/cm in the target. A fit
	of two Gaussians to the main peak 
	returns the position for the 122\,keV line at 2.54\,V.ns.}
      \label{fig:co57}
    \end{center}
  \end{figure}

  To convert the observed pulse energy (in mV) to electron
  equivalent energy (in keV) calibrations were performed with a
  $^{57}$Co (two spectral lines at 122\,keV and 136\,keV) source
  located beneath the target volume and delivered by a dedicated
  source delvivery mechanism. 10,000 events were recorded with
  waveforms 200$\mu$s in duration and with a drift field of
  1\,kV/cm. S1 and S2 signals were extracted from the parameterised
  data as described.  Figure~\ref{fig:co57} shows the
  calibration spectrum with a fit to the peak at 2.54\,V.ns. This
  corresponds to $\sim$67 pe (1\,pe $\approx$ 0.038\,V.ns) which gives
  a light-yield of $\sim$0.55\,pe/keV for this data. This corresponds
  to $\sim$ 1.1\,pe/keV at 0-field assuming a 50\% scintillation-yield loss
  with a drift field of 1\,kV/cm.

  \section{Summary}
  We have described the data acquisition system for the ZEPLIN\,II
  dark matter experiment which uses ACQIRIS digitizers to record the
  event waveforms from the detector. Approximately 8\,TB of dark matter
  data per year are recorded. Data reduction procedures have been
  developed to parameterise the data; up to 0.6\,TB of data per day
  can be reduced.
  
  \section{Acknowledgements}
  This work has been funded by the UK Particle Physics And Astronomy Re-
  search Council (PPARC), the US Department of Energy (grant number DE-
  FG03-91ER40662) and the US National Science Foundation (grant number
  PHY-0139065). We acknowledge support from the Central Laboratories for the
  Research Councils (CCLRC), the Engineering and Physical Sciences Research
  Council (EPSRC), the ILIAS integrating activity (Contract R113-CT-2004-
  506222), the INTAS programme (grant number 04-78-6744) and the Research
  Corporation (grant number RA0350). We also acknowledge support from Fun-
  dao para a Cincia e Tecnologia (pro ject POCI/FP/FNU/63446/2005), the
  Marie Curie International Reintegration Grant (grant number FP6-006651)
  and a PPARC PIPPS award (grant PP/D000742/1). We would like to grate-
  fully acknowledge the strong support of Cleveland Potash Ltd., the owners of
  the Boulby mine, and J. Mulholland and L. Yeoman, the underground facil-
  ity staff.
  
  
  \newpage
  \begin{table}[h]
    \begin{center}
      \caption{Waveform parameters calculated by reduction procedure.}
      \label{table:parameters}
      \begin{tabular}{|p{1in}|p{10cm}|}
	\hline
	    {\bf Parameter} & {\bf Explanation}\\
	    \hline
	    $B$ & Baseline level for each channel\\
	    $N_{RMS}$ & RMS noise level for each channel\\
	    $t_p$ & Start time of all pulses on each waveform\\
	    $t_0$ & Time of arrival of 10\% of the pulse charge\\
	    $w$& Width of each pulse in each channel\\
	    $A$& Area of each pulse in each channel\\
	    $h$& Height of each pulse in each channel\\
	    $FWHM$ & Pulse FWHM measured from maximum pulse
	    height\\
	    $LFWHM$ & Pulse FWHM measured from pulse start and
	    pulse end time\\
	    $\tau$& Charge mean arrival time\\
	    $t_v$ & Start time of veto pulse\\
	    $h_v$ & height of veto pulse\\
	    $A_v$ & Integrated area of veto pulse\\
	    \hline
      \end{tabular}
    \end{center}
  \end{table}
  \newpage
  \newpage
  \newpage
  \newpage
  \newpage
  \newpage
  \newpage
  \newpage
  \newpage
  \newpage
  

\begin{thebibliography}{99}

    \bibitem{naiad} G.J.Alner et al., Phys. Lett. B 616 (2005) 17-24.
    \bibitem{zeplin1} G.J.Alner et al., Astropart. Phys. 23 (2005)
      444-462.
    \bibitem{davies} G.J. Davies et al., Phys. Lett. B 320 (1994) 395.
    \bibitem{wang} H. Wang, PhD Thesis, UCLA, 1999.
    \bibitem{cline} D.B. Cline et al., Astropart. Phys. 12 (2000) 373.
    \bibitem{lewinsmith} J.D.Lewin and P.F.Smith, Astropart. Phys. 6
      (1996) 87-112.
    \bibitem{z2science} G.J.Alner et al. (2007), {\it Submitted to 
      Astroparticle Physics}
    \bibitem{polycold} \texttt{www.polycold.com}
      \bibitem{saes} \texttt{www.saesgetters.com}
    \bibitem{acqiris} \texttt{www.acqiris.com}
    \bibitem{datascan} \texttt{www.msl-datascan.com}
    \bibitem{knoll} G. F. Knoll, Radiation Detection and
      Measurement, third edition. Wiley
    \bibitem{adic} \texttt{www.adic.com}
    \bibitem{hbook} \texttt{wwwasdoc.web.cern.ch/wwwasdoc/
      hbook\_html3/hboomain.html}
  \end{thebibliography}
\end{document}